# Nonmonotonic Magnetic Friction from Collective Rotor Dynamics

**Hongri Gu[1,2]\*, Anton Lüders[1,3]\*, Clemens Bechinger[1]**

1. Department of Physics, University of Konstanz, 78464 Konstanz, Germany

2. Division of Integrative System and Design, Hong Kong University of Science and Technology, Hong Kong SAR, China

3. Department of Theoretical Physics, Universität Innsbruck, 6020 Innsbruck, Austria

\*these authors contributed equally to this work

**Amontons' law postulates a monotonic relationship between frictional force and the normal load applied to a sliding contact. This empirical rule, however, fails in systems where internal degrees of freedom — such as structural or electronic order — play a central role. Here, we demonstrate that friction can emerge entirely from magnetically driven configurational dynamics in the absence of physical contact. Using a two-dimensional array of rotatable magnetic dipoles sliding over a commensurate magnetic substrate, we observe a pronounced non-monotonic dependence of friction on the interlayer separation, and thus on the effective load. The friction peaks at an intermediate distance where competing ferromagnetic and antiferromagnetic interactions induce dynamical frustration and hysteretic torque cycles during sliding. Molecular dynamics simulations and a simplified two-sublattice model confirm that energy dissipation is governed by collective magnetic reorientations and their hysteresis. Our results establish the occurrence of sliding-induced changes in collective magnetic order, which has a strong impact on friction, and thus open new possibilities for contactless friction control, magnetic sensing, and the design of reconfigurable, wear-free frictional interfaces and metamaterials.**

## Introduction

Amontons' law, a time-honored and foundational empirical rule in tribology, posits a monotonic relationship between frictional force and normal load[1]. While widely applicable, this principle offers limited insight into the microscopic mechanisms underlying friction. Recent studies have revealed significant deviations from this law in systems where internal degrees of freedom — such as structural, electronic, or magnetic order — play a critical role[2–7]. These exceptions, often linked to phase transitions or symmetry breaking at sliding interfaces[8–10], underscore the potential of friction as a sensitive probe of interfacial dynamics. Magnetic contributions to friction have gained growing attention, particularly in contactless systems where dissipation is governed by configurational dynamics rather than mechanical abrasion[11–13]. Previous theoretical and numerical work has established a connection between magnetic ordering and friction[14–27], further supported by experiments using magnetic atomic force microscopy[11,28]. However, despite the atomic force microscopy's ability to detect magnetic interactions, it lacks the spatial resolution and sensitivity to fully resolve dynamic excitations at a microscopic spin level that govern friction at the nanoscale. Consequently, direct experimental evidence for the connection between spin dynamics and sliding friction is lacking.

In this study, we overcome this limitation by employing a two-dimensional array of rotatable magnetic dipoles that interact with a commensurate magnetic substrate. This macroscopic yet spatially resolved system allows us to directly track changes of collective magnetic configurations during sliding and correlate them with frictional behavior. Strikingly, we observe a non-monotonic dependence of friction on interlayer separation, with maximum dissipation occurring in a regime of frustrated magnetic alignment between ferromagnetic and antiferromagnetic order. This anomalous peak reveals a frictional mechanism rooted in sliding-induced magnetic hysteresis, absent of any mechanical contact. Our results demonstrate that magnetic friction can arise purely from internal reorientation dynamics and that frictional forces can serve as a sensitive and direct measure of magnetic order — accessible beyond the limitations of conventional techniques. These findings open pathways for designing tunable, wear-free frictional metamaterials[29–31] and advancing magnetic sensing technologies based on configurational control[32].

**Experimental setup**

We investigate the friction between two parallel layers in the $xy$-plane, each consisting of periodically arranged, millimeter-sized magnets in a square lattice with the same lattice constant $P$. The top layer (slider) comprises $7 \times 7$ neodymium-iron-boron (NdFeB) ring magnets with an outer diameter of $4\,\text{mm}$, an inner diameter of $1.2\,\text{mm}$, and a height of $4\,\text{mm}$ (Fig. 1a). Each magnet is mounted on a non-magnetic metal axis, allowing its magnetic moment $\vec{m}_{ij}$ to freely rotate within the $xz$-plane, forming an angle $\theta_{ij}$ relative to the $x$-axis (Fig. 1a). Here, the indices $i$ and $j$ denote the position of each rotor within the lattice. The entire rotor array is mounted on an aluminum frame attached to a three-axis resistive force sensor with a relative force resolution of approximately $0.02\,\text{N}$. More details of the force sensor are given in Sec. S1 of the Supporting Information (SI).

The bottom layer (substrate) is positioned parallel to the slider and consists of a $16 \times 9$ square lattice of cylindrical magnets. These magnets, made of the same NdFeB material, have a height and diameter of $4\,\text{mm}$ (Fig. 1b). Unlike the slider, the magnetic moments of the substrate magnets are fixed and oriented along the $x$-axis. The slider is translated relative to the substrate along the $x$-direction using a motorized stage at a constant velocity of $v = 8\,\text{mm/s}$, ensuring quasi-static conditions where the magnetic moments in the top layer effectively remain in equilibrium at each position during the sliding process.

To measure the orientations $\theta_{ij}$ of the rotors, the magnets are labeled with colored markers, which are tracked using a color-sensitive overhead camera (Fig. 1c). The vertical separation $h$ between the layers (height of the slider) is maintained using brass rollers mounted on the metal frame of the slider, which compensate for the magnetic attraction between the layers. Variations in $h$ are achieved by inserting spacers of different thicknesses between the rollers and the substrate. At the start of each experiment, the two layers are aligned parallel, with their magnetic lattices perfectly superimposed at the initial position. The slider is then translated forward and backward by seven lattice periods ($\Delta x_{\max} = 7 \times 16\,\text{mm}$) along the $x$-direction. During the translation, we record the $x$-component of the total force $F_x$ exerted on the slider using the force sensor, as well as the orientations $\theta_{ij}$ of the rotors. A typical example measurement of $F_x$ and a snapshot of the rotor orientations $\theta_{ij}$ during sliding are shown in Fig. 1c. Additional technical details of the setup components are provided in the Methods section.

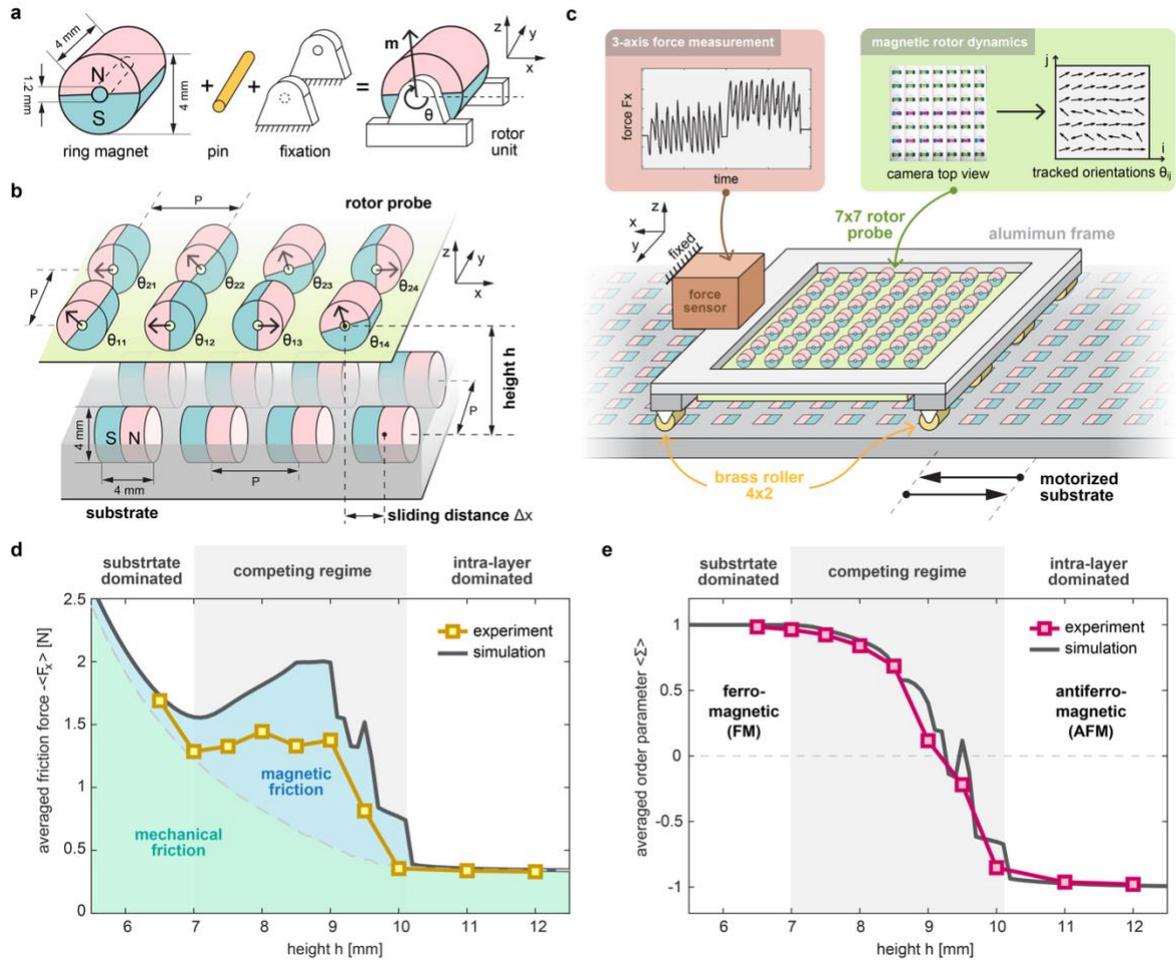

**Fig. 1 | Experimental setup, total friction, and order parameter.** (a) Magnetic rotor unit design: Ring-shaped NdFeB magnet mounted on a metal shaft enabling rotation around the $y$-axis with the orientation of the magnetic moment characterised by the angle $\theta$. (b) Illustration of the two interacting magnetic layers at separation (height) $h$ with square lattice geometry (lattice constant $P = $16mm). The top layer (7x7 magnets) is composed of magnetic rotors while the magnetic moments in the bottom layer (16x9 magnets) are fixed in $y$-direction. The relative horizontal displacement between the layers during sliding is characterized by $\Delta x$. (c) Mechanical structure for the sliding friction experiments: The setup includes a motorized stage, brass rollers for maintaining consistent separation, and a force sensor for direct measurement of the friction. The orientations of the rotors are directly observed using an overhead camera. (d) Total lateral friction $\langle F_x \rangle$ as a function of the height $h$ of the slider. The magnitude $-\langle F_x \rangle$ is presented, since $\langle F_x \rangle$ is negative. (e) Averaged orientation correlation order parameter $\langle \Sigma \rangle$ shown as a function of height $h$.

## Total friction and average magnetic order

Figure 1d shows how the measured mean sliding force $\langle F_x \rangle$ (the brackets correspond to the average over an integer number of lattice spacings $P$) varies as a function of the slider height $h$. This quantity is the total friction corresponding to the sliding process. Note that, while qualitatively consistent, the measured friction can strongly depend on small vertical misalignments. More information on its accuracy is given in Sec. S2.

Although the magnetic interaction, i.e. the normal load between the two layers monotonically decreases with increasing $h$, $\langle F_x \rangle$ exhibits a clear peak around $h_0 = 9.0\,\mathrm{mm}$. This is in strong contradiction with Amonton's empirical first law, which claims that sliding friction is proportional to the load[33]. The effective load due to magnetic attraction and the resulting differential friction coefficient for the setup are presented in Sec. S3.

The key to understanding this unusual friction behavior lies in analyzing how the magnetic order within the top layer evolves during the sliding process, as theoretical predictions suggest that changes in magnetization and magnetic excitations can couple to magnetic interaction-based friction — and *vice versa*[17,19,21,22,34,35]. To characterise the "relative" orientations of neighboring magnetic moments, we introduce the displacement-dependent order parameter

$$\Sigma(\Delta x) = \frac{1}{Z} \sum_{i=1}^{7} \sum_{j=1}^{6} \cos(\theta_{ij} - \theta_{ij+1}) \tag{1}$$

where $Z = 7 \times 6$ is a normalization factor. Accordingly, $\Sigma$ varies between +1 and -1. Here, +1 corresponds to a state in which all magnetic moments are aligned parallel to each other, representing a ferromagnetic (FM) order, whereas -1 corresponds to a state where neighboring moments regarding the $y$-direction are aligned anti-parallel, indicating an antiferromagnetic (AFM) order. When plotting the mean value $\langle \Sigma \rangle$ of the order parameter (averaged over an integer number of lattice spacings) vs. the layer distance $h$, we find a gradual change from FM to AFM ordering. Notably, $\langle \Sigma \rangle = 0$ (i.e., no average magnetic order) is located exactly at that distance where the peak in the friction force is observed (Fig. 1e). Similar friction peaks near structural rearrangements have been theoretically predicted[8] and experimentally measured[9,36] for materials externally heated close to a phase transition. This suggests a strong connection between the frictional behavior and changes of magnetic ordering, which will be discussed in the following.

**Competing intra- and interlayer magnetic interactions and layer interaction force**

Based on the $h$-dependent mean magnetic order parameter, we categorize the system into three distinct regimes with (i) ferromagnetic (FM) $\langle \Sigma \rangle \geq 0.95$, (ii) competing (CP) $-0.70 < \langle \Sigma \rangle < 0.95$ and (iii) antiferromagnetic (AFM) $\langle \Sigma \rangle \leq -0.70$ ordering, respectively. Figure 2a compares how the relative orientation of magnetic moments change during the sliding in the different regimes. To illustrate the differences in more detail, we discuss the orientational changes of two neighboring moments marked in blue and red located at $(i,j) = (4,3)$ and $(4,4)$, during the sliding process, respectively (Fig. 2b).

For small $h$, the magnetic field of the bottom layer is the dominant influence on the rotors, resulting in their parallel alignment. Since all rotors are exposed to the same magnetic field due to the commensurate conditions, this eventually leads to collective rotation during sliding of the rotors and thus to FM ordering. In contrast, at large $h$, the influence of the bottom layer weakens and the interactions between neighboring rotors prevail. This leads to AFM order with an alternating rotor orientation along the $y$-axis while being parallel along the $x$-axis (Fig. 2a, right column). During sliding, only a small collective "wiggling" of the rotors is observed (Fig. 2b, right column). At intermediate distances, where magnetic inter- and intra-layer interactions are comparable, the rotors periodically alternate between FM and AFM order in a rather discontinuous fashion (middle column in Fig. 2b). The dynamics of the moments corresponding to the three different regimes are presented in Video 1.

To corroborate our experiments, we additionally performed Molecular Dynamics (MD) simulations, where the magnets are modelled as point dipoles with their configuration and rotational degrees of freedom exactly matching the conditions in our experiments. The magnetic dipole strength has been determined from the magnetization of NdFeB provided by the manufacturer. To allow rapid relaxation of the dipoles and to maintain quasi-equilibrium conditions during the sliding process, we added an orientational friction term to the equations describing the angular motion of the magnetic moments. For more details, see the Methods section.

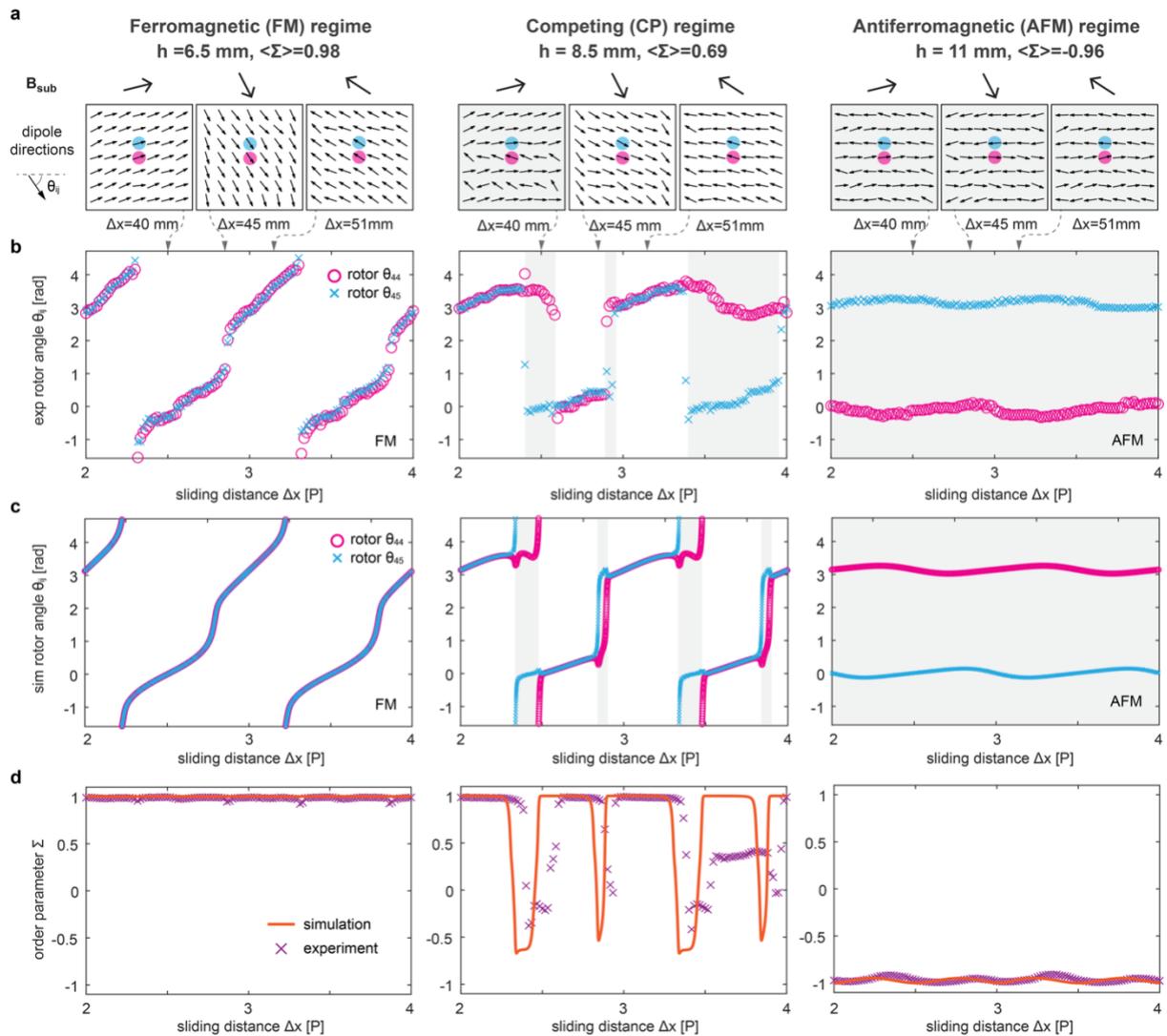

**Figure 2 | Magnetic moment configurations and dynamic responses at different heights $h$.** The orientation angle $\theta_{ij}$ of two example moments are depicted as a function of the displacement $\Delta x$ for three selected heights $h = 6.5\,\mathrm{mm}$ (left), $8.5\,\mathrm{mm}$ (middle), and $11\,\mathrm{mm}$ (right) representing the different dynamical states (FM, CP, AFM). (a) Experimental snapshots of the $7 \times 7$ rotor array configurations at different displacements $\Delta x$. The local magnetic field acted by the substrate on the moments is indicated above each snapshot. (b, c) Experimental and simulation results for the tagged $\theta_{ij}$ showing the displacement-dependent dynamic

response of two selected rotors from the array at each height. (d) Orientation correlation order parameter $\Sigma$ as a function of the Displacement $\Delta x$.

Figure 2c shows the corresponding results of the simulations matching the experimental data in Fig. 2b. In the FM and AFM regime, the dynamics of the experiments and the simulations are in good agreement. In the CP regime, the jumps between parallel and antiparallel configurations appear less periodic in the experiments, which is due to some small variations in the purchased magnets and small deviation in the structure dimensions limited by the printing resolution of the 3D printer used to manufacture the base of the slider (Methods). For a similar reason, the flipping of magnetic moments between parallel and antiparallel direction is not perfectly reproduced in the simulations because the system is highly unstable at the tipping point and thus susceptible to small perturbations. However, both the simulations and the experiments exhibit the dynamical transition between the FM and AFM ordering in the CP regime. A direct comparison between the experiments and the simulations are shown in Video 2.

In Fig. 2d, we compare the displacement-dependent order parameter $\Sigma(\Delta x)$ obtained from experiments and simulations. Consistent with above, the orientation correlation of magnetic dipoles hardly changes in the FM and AFM regime but strongly alternates in the CP regime. Such variations are also responsible for the small value of $\langle\Sigma\rangle$ at intermediate distances (Fig. 1e). The full height dependence of $\langle\Sigma\rangle$ obtained from simulations is shown in Fig. 1e as a solid line and fits excellently to the experimental data. The detailed dynamics of the magnetic moments during the transition between the three regimes is presented in Video 3.

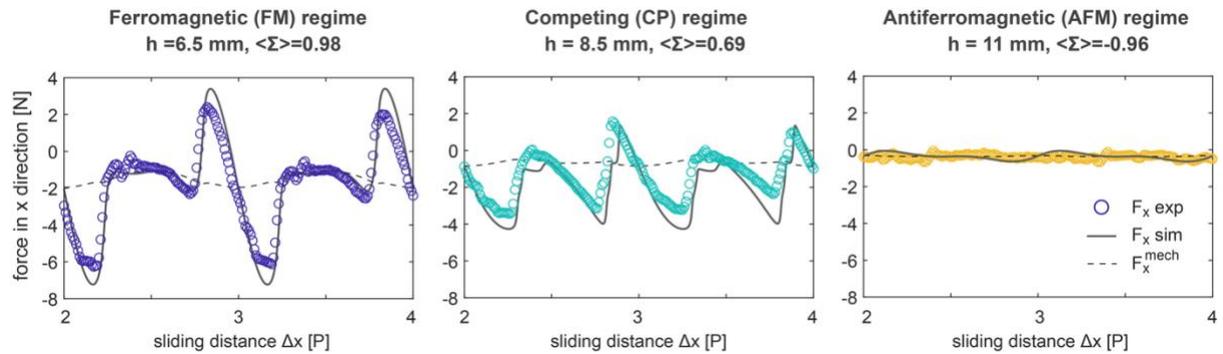

**Figure 3 | Height-dependent magnetic interaction force between the two layers.** Total force acting on the top layer as a function of the displacement $\Delta x$ obtained from the force sensor (symbols) and the simulations (lines). The left, middle, and right columns represent the FM ($h = 6.5\,\mathrm{mm}$), the CP ($h = 8.5\,\mathrm{mm}$), and the AFM regimes ($h = 11\,\mathrm{mm}$), respectively. The dashed lines are the mechanical contributions $F_x^{\mathrm{mech}}(\Delta x)$ of the friction obtained from the simulations.

Figure 3 shows the experimentally measured forces $F_x$ (symbols) during the sliding process for the three different regimes with heights $h = 6.5\,\mathrm{mm}$ (FM), $8.5\,\mathrm{mm}$ (CP), and $h = 11\,\mathrm{mm}$ (AFM), respectively. Notably, $F_x$ contains both, magnetic $F_x^{\mathrm{mag}}$ and mechanical $F_x^{\mathrm{mech}}$ friction forces, where the latter is mostly caused by the brass rollers compensating for the magnetic attraction (load) between the two layers. Since we are mainly interested in the magnetic friction, both contributions need to be separated, which is discussed in the following.

The mechanical part $F_x^{\mathrm{mech}}$ contains two contributions $F_1^{\mathrm{mech}}$ and $F_2^{\mathrm{mech}}$. The first part is due to the magnetic attraction $F_z^{\mathrm{mag}}$ between the layers leading to contact friction due to the brass rollers. Using Amanton's law, we approximate $F_1^{\mathrm{mech}} = \mu F_z^{\mathrm{mag}}$ with the friction coefficient $\mu = 0.09$ being experimentally determined by variation of the load between the layers (Sec. S4). In addition, another small load-independent source of friction exists which is likely caused by a small error in the initial zero-load conditions caused by vertical misalignment when calibrating the setup (see also Sec. S2). This can be seen for largest distances (i.e., $h = 11.0\,\mathrm{mm}$) in Fig. 1c and Fig. 3, where $F_z^{\mathrm{mag}} \approx 0$ and therefore $F_1^{\mathrm{mech}} \approx 0$. The small deviation from $\langle F_x \rangle = 0$ and $F_x = 0$ is here attributed to $F_2^{\mathrm{mech}} \approx -0.34\,\mathrm{N}$. More details on the mechanical friction model can be found in Sec. S4.

The solid lines in Fig.3 correspond to simulations where the above mechanical contributions have been taken into account. Despite the simple dipole approximation, the simulations are in excellent agreement with the experiments. In the FM regime, we observe large oscillations in $F_x(\Delta x)$ due to the strong interaction between the slider and the bottom layer. Such fluctuations gradually decrease with decreasing magnetic interaction strength and essentially vanish for $h = 11\,\mathrm{mm}$. Notably, in the FM and AFM case, $F_x$ almost symmetrically fluctuates around the mechanical contribution $F_x^{\mathrm{mech}}$ (marked by the dashed lines in Fig. 3 and approximated from the simulation data). This suggests that the mean magnetic friction $\langle F_x^{\mathrm{mag}} \rangle$ (averaged over an integer number of lattice periods) almost vanishes and only $\langle F_x^{\mathrm{mech}} \rangle$ contributes to the total friction $\langle F_x \rangle$. This is in strong contrast to the CP regime where $\langle F_x^{\mathrm{mag}} \rangle$ clearly deviates from zero, which results in large magnetic friction.

**Non-monotonic magnetic friction and its relationship to hysteresis**

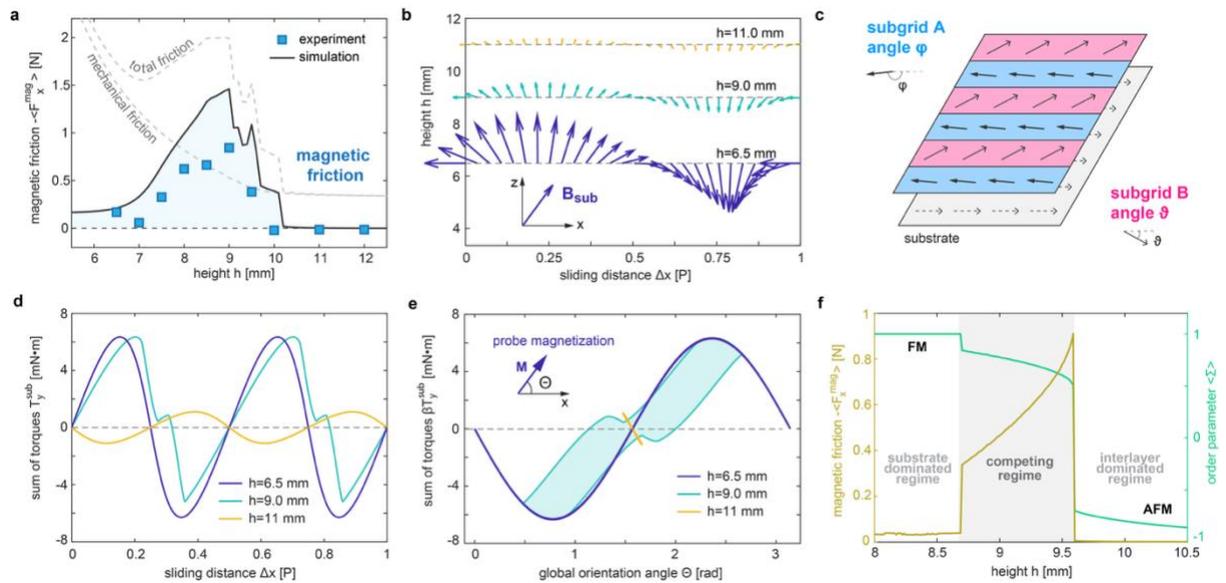

**Figure 4 | Non-monotonic magnetic friction and its underlying mechanism.** (a) Magnetic friction $\langle F_x^{\mathrm{mag}} \rangle$ in dependence of the height $h$. The additional subpeaks and plateaus of the simulation results are finite size effects, where the moments on the edges rotate differently than the rest of the moments. (b) Visualization of the magnetic substrate field $\vec{B}_{\mathrm{sub}}$ in dependence of the displacement $\Delta x$ and the height $h$ obtained from the simulations. (c) Schematic sketch of the simplified model, where the slider is divided in two sublattices. (d)

Sum of torques $T_y^{\text{sub}}$ acted by the substrate field on the top layer obtained from the simplified model. (e) Sim of torques $T_y^{\text{sub}}$ plotted over the angle $\Theta$ with $|\vec{M}|\cos\Theta = \hat{\vec{x}} \cdot \vec{M}$ between the movement direction $\hat{\vec{x}}$ and the magnetization $\vec{M}$ of the top layer (see inset) to highlight the hysteresis. To depict the correct area that describes the dissipated energy, $T_y^{\text{sub}}$ is scaled with the sign $\beta$, which is positive when $\Theta(\Delta x)$ decreases and negative when $\Theta(\Delta x)$ increases (see Sec. S11). (f) Magnetic friction $\langle F_x^{\text{mag}} \rangle$ and average order parameter $\langle \Sigma \rangle$ calculated from the simplified model.

Subtracting the mechanical friction from the sensor measurements allows us now to isolate the magnetic friction force. Figure 4a shows the experimental (symbols) and numerical results (lines) of $\langle F_x^{\text{mag}} \rangle$ as a function of $h$. Although the magnetic attraction, i.e. the load on the bottom layer, continuously decreases with increasing distance $h$ (see Sec. S3), the friction displays a maximum at $h_0 \approx 9.0\,\text{mm}$ (CP regime). Such behaviour is in strong contrast to the monotonic load-dependence of mechanical friction (Amaton's law, dashed line). The magnetic friction peak can also be successfully reproduced by reconstructing the experimental friction using the measured orientations $\theta_{ij}(\Delta x)$ and calculating $\langle F_x^{\text{mag}} \rangle$ under the assumption of magnetic dipole interactions (see Section S5). This confirms that the observed peak in magnetic friction is independent of the applied model for the mechanical contribution of the friction.

To understand the unusual load-dependence on magnetic friction, we consider the Hamiltonian

$$\mathcal{H} = \mathcal{H}_0 + \mathcal{H}_{\text{sub}} + \mathcal{H}_{0,\text{sub}} \tag{2}$$

of an idealized system, where $F_x^{\text{mech}} \equiv 0$. Here, $\mathcal{H}_0$ is the total energy of the slider, $\mathcal{H}_{\text{sub}}$ denotes the magnetic exchange energy between the layers, and $\mathcal{H}_{0,\text{sub}}$ the internal magnetic energy of the bottom layer (being constant). The energy transfer from and to the top layer during sliding is then given by the time derivative of $\mathcal{H}_0$. Using the rotors' equations of motion, we find (see Sec. S6 for the derivation)

$$\frac{d\mathcal{H}_0}{dt} = \sum_{i,j} \left[ f(\dot{\theta}_{ij})\dot{\theta}_{ij} - \frac{\partial \mathcal{H}_{\text{sub}}}{\partial \theta_{ij}}\dot{\theta}_{ij} \right]. \tag{3}$$

The first term on the r.h.s. accounts for the energy dissipation rate owing to the rotor's mechanical friction within their shafts, where $f(\dot{\theta}_{ij})$ is the microscopic shaft friction. The second term considers the power required for rotation of the magnets in presence of the substrate field $\vec{B}_{\text{sub}}$. Because the $\vec{B}_{\text{sub}}$ rotates as a function of $\Delta x$ (see Fig. 4b), it exerts torques $\vec{T}_{ij}^{\text{sub}} = \vec{m}_{ij} \times \vec{B}_{\text{sub}}$ on the magnetic moments $\vec{m}_{ij}$. The torque components parallel to the rotating axis $[\vec{T}_{ij}^{\text{sub}}]_y = -\partial \mathcal{H}_{\text{sub}}/\partial \theta_{ij}$ leads to the rotation of the moments as experimentally observed (Fig. 2a/b). To simplify Eq. (3), we use of the commensurability between both layers, which renders $\mathcal{H}_0$ periodic with respect to the lattice constant $P$. Accordingly, when taking the average over $P$ and considering a steady state, $\langle d\mathcal{H}_0/dt \rangle = 0$, and we find

$$\sum_{i,j}\langle f(\dot{\theta}_{ij})\dot{\theta}_{ij}\rangle = \sum_{i,j}\langle \frac{\partial \mathcal{H}_{\text{sub}}}{\partial \theta_{ij}}\dot{\theta}_{ij}\rangle = -\sum_{i,j}\langle [\vec{T}_{ij}^{\text{sub}}]_y \dot{\theta}_{ij}\rangle. \tag{4}$$

This implies that the averaged energy dissipation rate (l.h.s) is only given by $\vec{T}_{ij}^{\text{sub}}$ and $\dot{\theta}_{ij}$ (r.h.s). We will discuss these two parameters below.

The experimental observation that the magnetic moments of the slider exhibit within good approximation always either ferromagnetic or antiferromagnetic order (see Fig. 2a/b) allows for a simplified model, where the moments of the slider are arranged within two sublattices with different magnetic orientations (Fig. 4c). Because the moments within each sublattice are always parallel, the top layer is fully characterized by two rotational degrees of freedom $\varphi$ and $\vartheta$. Considering the overdamped limit where the slider energy $\mathcal{H}_0$ is only determined by magnetic interactions $\mathcal{H}_I$ within the slider and applying only nearest neighbor interactions, $\mathcal{H}_0$ can be splitted into the interaction energy of moments within the same sublattice

$$\mathcal{H}_{\text{self}} = a\sin^2(\varphi) + a\sin^2(\vartheta) \tag{5}$$

and the magnetic exchange energy across the two sublattices

$$\mathcal{H}_{\text{int}} = b\cos(\varphi - \vartheta), \tag{6}$$

with $\mathcal{H}_0 = \mathcal{H}_{\text{self}} + \mathcal{H}_{\text{int}}$. The full derivation is detailed in Sec. S7. The amplitudes $a$ and $b$ are derived for the geometry of our setup and matched to the specifications of the magnets. Because the substrate magnetic field corresponds to a rotating field in the comoving frame of the top layer, we approximate

$$\mathcal{H}_{\text{sub}} = c(h)\cos(\varphi - k\Delta x) + c(h)\cos(\vartheta - k\Delta x) \tag{7}$$

with $k = 2\pi/P$ and interaction strength $c(h) \sim 1/h^3$, again matched to the interactions in the experiments. Assuming $f(\dot{\varphi}) \approx -\Gamma\dot{\varphi}$ and $f(\dot{\vartheta}) \approx -\Gamma\dot{\vartheta}$ with $\Gamma$ a generalized friction coefficient associated with the microscopic change of the magnetization, we arrive at the equations of motion $\Gamma\dot{\varphi} = -\partial\mathcal{H}/\partial\varphi$ and $\Gamma\dot{\vartheta} = -\partial\mathcal{H}/\partial\vartheta$, which can be solved numerically (Methods). The corresponding total dissipated energy can be obtained by integrating Eq. (4), which can then be used to calculate the friction. Details are given in Sec. S8.

Fig. 4d shows how the sum of torques $T_y^{\text{sub}} = -\partial\mathcal{H}_{\text{sub}}/\partial\varphi - \partial\mathcal{H}_{\text{sub}}/\partial\vartheta$ varies for different heights. For $h = 6.5\,\text{mm}$ the torque oscillates smoothly which leads to an approximately constant angular velocity $\omega \approx kv$ as experimentally confirmed in Fig. 2a/b (for $h \to 0$, $\omega = kv$ becomes perfectly constant, see Sec. S9). Because $T_y^{\text{sub}}$ alternates symmetrically around $T_y^{\text{sub}} = 0$ during one lattice constant, we obtain from Eq. (4) $\langle\partial\mathcal{H}_{\text{sub}}/\partial\varphi\,\dot{\varphi}\rangle + \langle\partial\mathcal{H}_{\text{sub}}/\partial\vartheta\,\dot{\vartheta}\rangle \approx -\langle T_y^{\text{sub}}\rangle\omega \approx 0$ for the average energy dissipation rate, i.e. a vanishing magnetic friction (due to the mechanical shaft friction of the magnets, the friction is not exactly zero). For intermediate distances, $h = 9.0\,\text{mm}$, however, $T_y^{\text{sub}}$ no longer symmetrically oscillates around $T_y^{\text{sub}} = 0$, which results in a strong magnetic friction as observed in our experiments. For $h = 11.0\,\text{mm}$, the torque is small while the rotational velocity of the moments is close to zero (c.f. Fig. 2 a/b). This leads to a vanishing magnetic friction for large $h$, and, thus, explains the experimentally observed maximum in the magnetic friction $\langle F_x^{\text{mag}}\rangle$ as a function of $h$ (Fig. 4a). Analytic approximations for $\langle F_x^{\text{mag}}\rangle$ based reflecting the discussed behavior of the different regimes, are presented in Sec. S9.

The asymmetric variation of the torque as a function of $\Delta x$ leads to the signature of a magnetic hysteresis behavior similar to that observed in a ferromagnet (see Sec. S10). Such a behavior was also reported earlier for magnetic friction measurements[28], as well as in simulations on energy dissipation in Ising layers[14,37,38]. The observed hysteresis is shown in Fig. 4e, where we plotted $T_y^{\text{sub}}$ obtained from numerical integration of the simplified model as a function of the angle $\Theta$ between the slider's magnetization $\vec{M} = \sum_{ij} \vec{m}_{ij}$ and the $x$-axis (inset Fig. 4d). Within the chosen representation, the area within the hysteresis loops corresponds to the dissipated energy during sliding (see Sec. S11 for a derivation), showing the same nonlinear dependence on $h$. The signature of the hysteresis can also be understood from the simplified Hamiltonian of our system, which we discuss in Sec. S12.

Calculating $\langle F_x^{\text{mag}} \rangle$ from our simplified model (Methods), the result is indeed in excellent qualitative agreement with our results thus far (Fig. 4f). Notably, our simplified model also confirms the observed connection between $\langle F_x^{\text{mag}} \rangle$ and the averaged order parameter $\langle \Sigma \rangle$ (Fig. 4f). This underlines the direct link between the measured changes in the magnetic arrangements and the resulting friction. Using the MD simulations, we can confirm all our findings based on the simplified model for the the full system taking all $7 \times 7$ rotational degrees of freedom into account (see Sec. S13).

**Discussion and Outlook**

Our results demonstrate that magnetic friction can be precisely tuned by controlling the microscopic magnetic order at the interface. This aligns with theoretical predictions[8,14,34] and recent atomic force microscopy measurements[9,36], which suggest that friction increases sharply in systems near structural rearrangements or phase transitions, regardless of their origin. Unlike these previous studies, where such changes in order are externally initiated by temperature control or applied fields, the interactions during sliding themselves drive the transition within the studied magnetic layers, with peak dissipation arising from dynamic switching between ferromagnetic and antiferromagnetic states. The macroscopic nature of our setup provides unobstructed access to the structural dynamics. This offers unique insights into the interplay between sliding and structural order, while overcoming the limitations of atomic force microscopy-based techniques, which cannot directly monitor the dynamics of underlying phase transitions and excitations. Expanding our model system to incommensurate magnetic configurations and mismatching magnetic layers, this macroscopic nature of the setup could also be exploited to explore the dynamics and propagation of topological magnetic solitons[39–42], which are hard to access experimentally in conventional magnetic materials without modifying the underlying structure[43].

The observed breakdown of Amontons' law establishes friction as a sensitive probe of collective magnetic dynamics and paves the way for wear-free, tunable interfaces. Beyond macroscopic rotor arrays, the principles demonstrated here are broadly applicable to low-dimensional magnets, spintronic materials, and XY model-type systems with similar energetic symmetries, where frictional anomalies could signal or enable magnetic switching[27]. These findings open avenues for the design of friction metamaterials[29–31,44,45] — engineered surfaces that dissipate energy in a programmable manner. By tailoring magnetic hysteresis through carefully arranged micromagnets[32], such systems could enable reconfigurable, energy-efficient interfaces controlled by internal degrees of freedom.

**Acknowledgement**

We acknowledge stimulating discussions with Andrea Vanossi. We thank Dieter Barth for advice and technical assistance in building the force testing setup, Davide Bossini and Veit-Loretz Hueta for fruitful discussions, Laurence Carls for help with communications, and Thomas Franosch for his valuable input regarding the calculation of the energy transfer rates. The authors acknowledge support by the local computing resources through the core facility SCCKN.

**Methods**

*Experiments:* A single ring NdFeB magnet (N35 grade with nickel surface coating, HKCM Article No. 9963-73617) serves as the rotor, and a cylindrical magnet of the same kind (HKCM Article No. 9962-61814) serves as the substrate. It is radially magnetized so that its dipole moment can freely rotate about a fixed metal pin, providing one rotational degree of freedom. The metal pin is an 8 mm-long dowel, 1 mm in diameter, made of 18-8 stainless steel (McMaster-Carr part number 91585A062).

A commercial fused deposition modeling (FDM) 3D printer (Bambu Lab X1 Carbon) and standard PLA filament were used to fabricate the plastic probe, to which the pin was then affixed. This setup allows the cylindrical magnet to rotate around the pin while remaining attached to the probe.

To maintain a consistent gap between the probe and the substrate, we incorporated brass rollers that serve as direct mechanical supports during sliding. These rollers help stabilize the probe against the strong magnetic attraction between the 7x7 rotor array and the substrate. Each brass roller has a cylindrical form similar to the rotors, and additional 3D-printed spacers of varying heights ensure discrete probe-substrate separations. This arrangement enables smooth gliding of the probe across the substrate surface.

*Simulations:* To obtain numerical data for our setup, we integrate the dynamics of the angles $\theta_{ij}$ with the Verlet algorithm[46]

$$\theta_{ij}(t+\Delta t) = 2\theta_{ij}(t) - \theta_{ij}(t-\Delta t) + \frac{1}{I}f(\dot{\theta}_{ij}(t))\Delta t^2 + \frac{1}{I}T_{ij}^{\text{tot}}(t)\Delta t^2$$

using a time step of $\Delta t = 3 \times 10^{-4}\,\text{s}$. Here, $I$ is the moment of inertia of the rotors of the experiments, and $T_{ij}^{\text{tot}} = \vec{m}_{ij} \times \vec{B}$ is the magnetic torque acting on the moment $\vec{m}_{ij}$ due to the total magnetic field $\vec{B}$ generated by the moments of the substrate and the other moments on the slider. We assume dipole interaction, which means all moments $\vec{m}$ generate a field[47]

$$\vec{B}_{\text{dipole}}(\vec{r}) = \frac{\mu_0}{4\pi|\vec{r}|^3}\left[\frac{3(\vec{m}\cdot\vec{r})\vec{r}}{|\vec{r}|^2} + \vec{m}\right],$$

where $\mu_0$ is the vacuum permeability. The geometry and parameters of the simulations are matched to the experimental values. For the friction term in the integrator for $\theta_{ij}$, we assume $f(\dot{\theta}_{ij}(t)) = -\gamma\dot{\theta}_{ij}(t)$, where the friction coefficient $\gamma$ is a free parameter that is adjusted for stability and to match the experimental results. The total magnetic interaction force between the top and the bottom layer is calculated by summing over all dipole interaction forces

$\vec{F}_{ij} = \nabla(\vec{m}_{ij} \cdot \vec{B}_{\text{sub}})$ acting between moments of the slider and the substrate. More details are given in Sec. S14.

*Simplified model:* Calculating the partial derivatives of the Hamiltonian $\mathcal{H}$ of the simplified model regarding the two degrees of freedom $\varphi$ and $\vartheta$ and taking the shaft friction into account, we obtain two coupled, nonlinear differential equations. These equations are solved numerically using a standard solver based on the explicit Runge-Kutta (4,5) formula[48,49]. The dissipated energy $E_{\text{diss}}$ can be obtained by formulating Eq. (2) for the two remaining degrees of freedom, integrating the right-hand side regarding the time needed to translate one lattice constant $P$, and inserting the solutions for $\varphi$ and $\vartheta$ (see also Sec.S7 of the SI). The observables are calculated via $\langle F_x^{\text{mag}} \rangle = E_{\text{diss}}/P$ and $\langle \Sigma \rangle = \langle \cos(\varphi - \vartheta) \rangle$. More information regarding the numerical solution of the simplified model is given in Sec. S15.